\newcommand{\Figurebb}[9]{
\begin{figure}[H]\begin{center}
\leavevmode
\epsfysize=#7cm
\epsfbox[#2 #3 #4 #5]{#6}
\par
\parbox{#8cm}{
\caption[figure]{\renewcommand{\baselinestretch}{0.8} \small
                                           \hspace{-0.3truecm}#9}
\label{#1}}
\end{center}
\end{figure}
}

\def\fig#1{Fig.~\ref{#1}}

\def\bs{\bigskip}

\def\siml{\,\hbox{\kern.1em \lower.6ex \hbox{$\sim$} \kern-1.12em
          \raise.6ex \hbox{$<$} }}
\def\simg{\,\hbox{\kern.1em \lower.6ex \hbox{$\sim$} \kern-1.12em
          \raise.6ex \hbox{$>$} }}

\def\d3r{d^3r\;}

\def\eq#1{(\ref{#1})}

\def\be{\begin{equation}}
\def\ee{\end{equation}}
\def\bea{\begin{eqnarray}}
\def\eea{\end{eqnarray}}

\documentstyle[prl,aps,epsf,here]{revtex}
\twocolumn

\begin{document}

\draft
\preprint{TPR-00-13}
\title{Wavefunction localization and its semiclassical description\\
       in a 3-dimensional system with mixed classical dynamics}
\author{M. Brack$^a$, M. Sieber$^b$ and S. M. Reimann$^{c,a}$}
\address{
$^a$Institute for Theoretical Physics, Regensburg University,
    D-93040 Regensburg, Germany\\
$^b$Max-Planck-Institute for the Physics of
    Complex Systems, N\"othnitzer Str. 38, D-01187 Dresden, Germany\\
$^c$Lund Institute of Technology, P. O. Box 118, S-22100 Lund, Sweden
}

\date{\today}

\maketitle

\vglue -1.0truecm

\begin{abstract}

We discuss the localization of wavefunctions along planes containing
the shortest periodic orbits in a three-dimensional billiard system
with axial symmetry. This model mimicks the self-consistent mean field 
of a heavy nucleus at deformations that occur characteristically during 
the fission process \cite{fuhi,sven}. Many actinide nuclei become 
unstable against left-right asymmetric deformations, which results in 
asymmetric fragment mass distributions.
Recently we have shown \cite{fis1,fis2} that the onset of this
asymmetry can be explained in the semiclassical periodic orbit theory
by a few short periodic orbits lying in planes perpendicular to the
symmetry axis. Presently we show that these orbits are surrounded by
small islands of stability in an otherwise chaotic phase space, and
that the wavefunctions of the diabatic quantum states that are most
sensitive to the left-right asymmetry have their extrema in the same
planes. An EBK quantization of the classical motion near these planes
reproduces the exact eigenenergies of the diabatic quantum states
surprisingly well.

\end{abstract}


\narrowtext

We have recently applied \cite{fis1,fis2} the periodic orbit theory
\cite{gutz,bb} to a three-dimensional cavity model with axially
symmetric deformations that typically occur near the isomer minimum and
the second maximum of the characteristic double-humped fission barrier
\cite{fuhi,sven} of many actinide nuclei. The boundary of the cavity in
cylindrical coordinates $(z,\rho)$ is given by a shape function
$\rho(z)$; the deformations are defined by an elongation parameter $c$
and an octupole-type left-right asymmetry parameter $\alpha$ (see Ref.\
\cite{fuhi} for the detailed definitions; the neck parameter $h$ is
fixed to be zero in the present work). The shortest periodic orbits are
found in planes perpendicular to the symmetry ($z$) axis at locations
$z_i$ given \cite{bb} by $\rho'(z_i)=0$; they are just the shortest
polygons (diameter, triangle, square, etc.) inscribed into the circular
cross sections of the billiard system with radii $\rho(z_i)$. The
oscillating part $\delta E$ of the total energy of the system
containing $N$ fermions (we do not distinguish neutrons from protons
and neglect the Coulomb and spin-orbit interactions) was calculated by
the appropriate semiclassical trace formula. A uniform approximation
was introduced to describe the bifurcation of a single orbit plane (at
$z_0=0$) into three orbit planes $z_i$ ($i=0,1,2$) at the moment of the
neck formation where the shape function $\rho(z)$ becomes convex (see
Ref.\ \cite{fis1} for details). The gross-shell features in $\delta E$
were emphasized by convolution of the trace formula over the wave
number $k$ by a Gaussian of width $0.6/\!R$ ($R$ is the radius of the
spherical nucleus), producing a smearing of the shell structure similar
to that caused by the residual pairing interaction in realistic nuclear
models \cite{fuhi}.

In the quantum-mechanical calculations for the energy shell correction
$\delta E$, derived from the quantum spectrum of realistic nuclear
shell-model potentials using Strutinsky's shell-correction method
\cite{stru}, it was found that the outer fission barrier of many
actinide nuclei is unstable against octupole-type deformations
\cite{asym}. In our present parametrization, this barrier is located at
$c=1.53$; by varying $\alpha$ from zero to $\sim 0.13$, the energy
$\delta E$ is lowered by about $\sim 2.5 - 3$ MeV. (For the present
qualitative discussion, we neglect the smooth liquid-drop model part of
the total energy which varies much less in this restricted region of
deformations.) This result could be well reproduced in our
semiclassical calculations \cite{fis1,fis2}. Thereby, it

\Figurebb{pcss1}{40}{20}{568}{685}{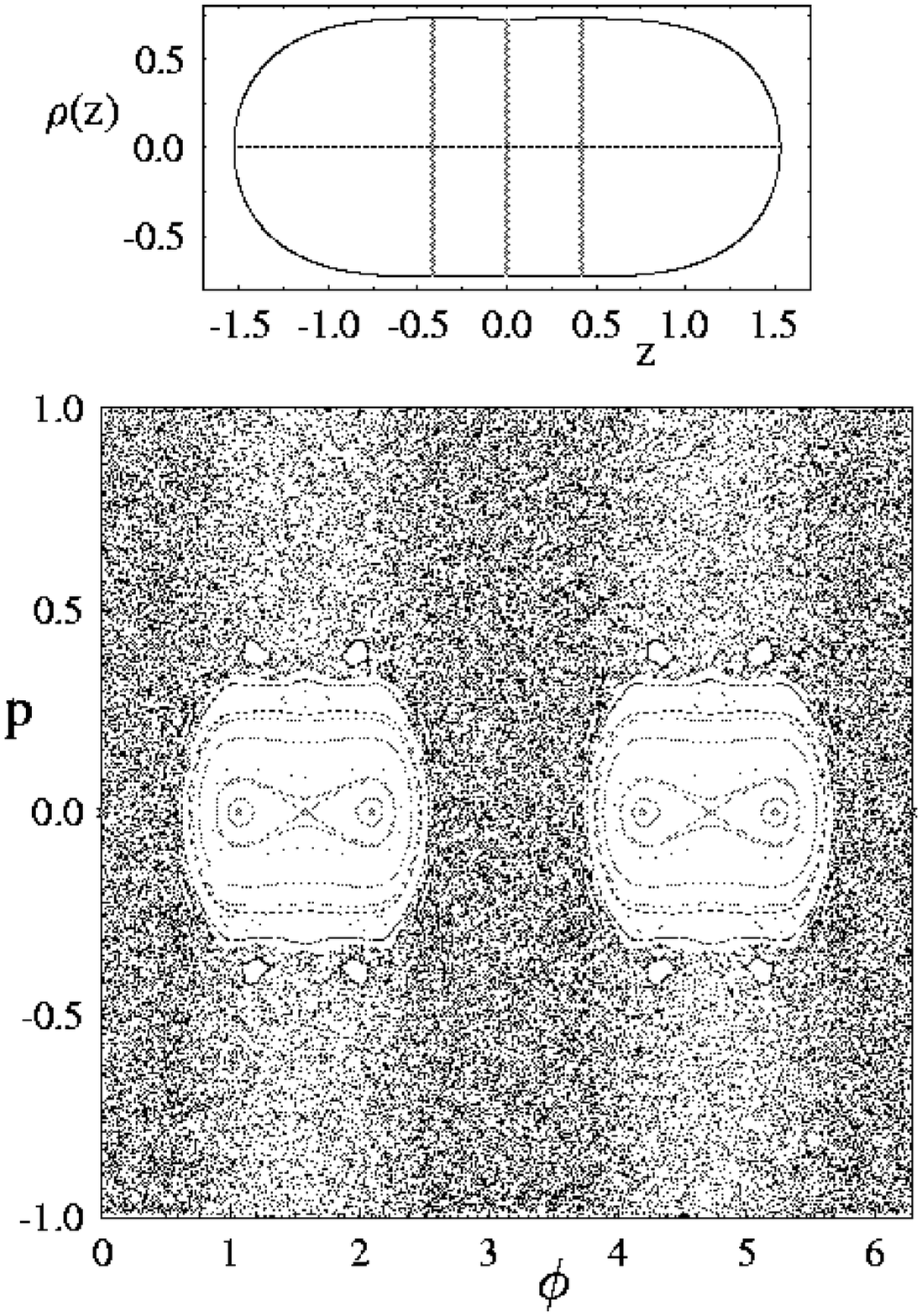}{10.28}{6.0}{
{\it Top:} Cavity shape $\rho(z)$ at elongation $c=1.53$ and asymmetry
$\alpha=0.0$. The vertical lines indicate the planes containing the
shortest periodic orbits. {\it Bottom:} Poincar\'e surface of section
$(p,\phi)$ for $L_z=0$. See the text for details.
}

\noindent
turned out to be sufficient to include the two shortest periodic orbits
(diameters and triangles) in each of the planes $z_i$ into the trace
formula; the contributions of longer orbits were shown to be negligible
\cite{fis2}.

In the upper parts of Figs.\ \ref{pcss1} and \ref{pcss2}, the shape
boundaries $\rho(z)$ are shown for two deformations with fixed
elongation $c=1.53$. The first corresponds to the symmetric outer
barrier with $\alpha=0$, and the second to the asymmetric saddle with
$\alpha=0.13$. The vertical lines show the positions $z_i$ of the
planes containing the periodic orbits. In the symmetric case
(\fig{pcss1}), the shortest orbits at $z_0=0$ are slightly unstable,
wheras those at $-z_2=z_1=0.414$ are stable. In the asymmetric case
(\fig{pcss2}), there is only one plane at $z_0=0.671$ containing stable
orbits. In the lower parts of the figures, we show Poincar\'e surfaces
of section $(p,\phi)$ for trajectories with (conserved) angular
momentum component $L_z=0$, where $p$ is the component of the momentum
parallel to the tangent plane at a reflection point and $\phi$ is the
polar angle in the $(z,\rho)$ plane of a reflection point at the
boundary. (Note that the Poincar\'e mapping in these variables is not
area preserving; this is, however, immaterial for the qualitative
interpretation of the figures.) In both cases we find two main islands
of stability containing the fixed points corresponding to the shortest
stable and unstable orbits mentioned above. Apart from some KAM chains
of smaller islands corresponding to higher resonances, the remainder
part of the phase space is mainly chaotic. Note that the overall
chaoticity is much larger in the energetically

\Figurebb{pcss2}{40}{20}{568}{740}{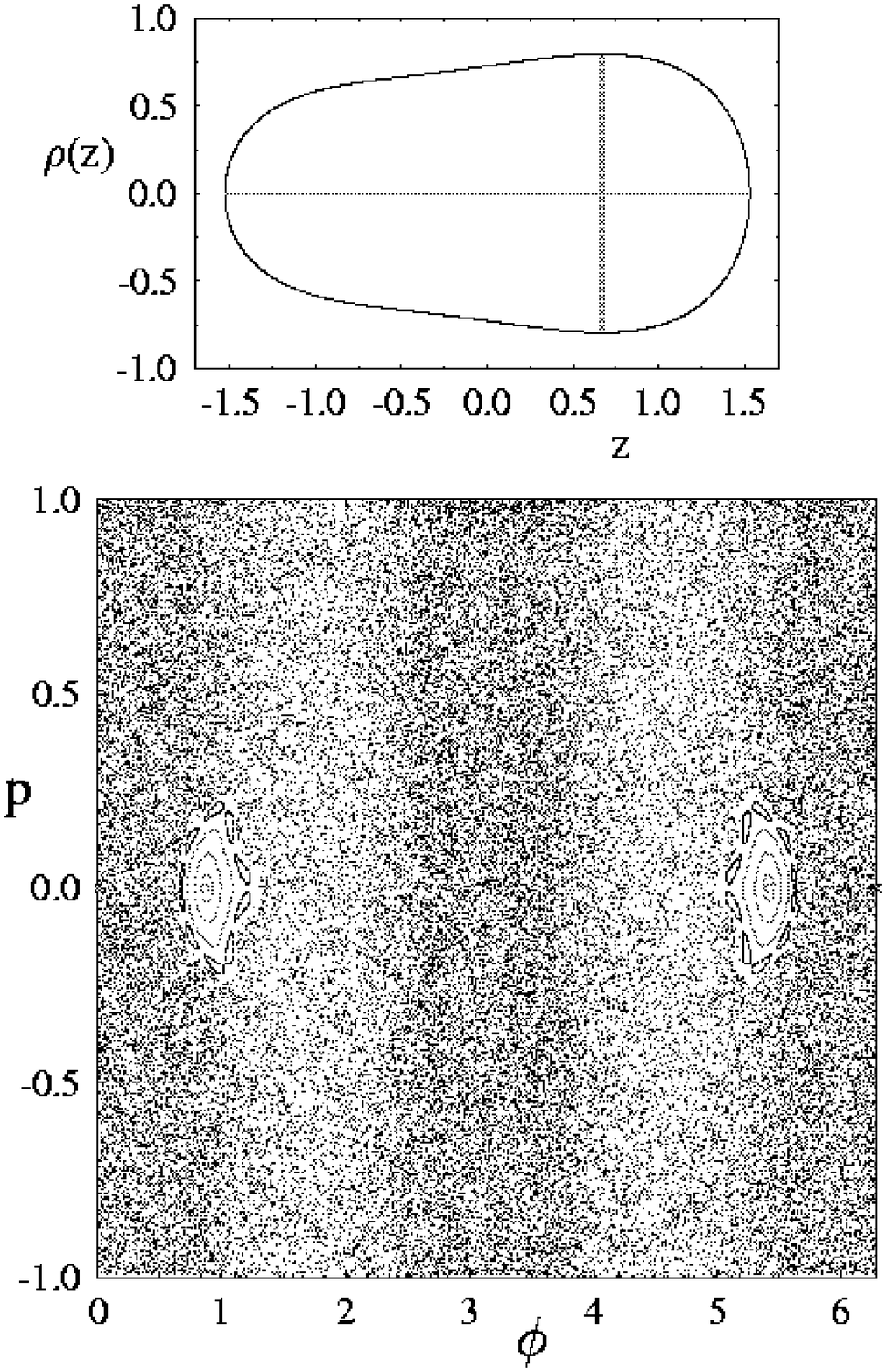}{10.42}{6.0}{
The same as in Fig.\ \protect\ref{pcss1} for asymmetry $\alpha=0.13$.
}

\noindent
more stable asymmetric case. We thus find that the shell effect which
leads to the energetic instability of the symmetric outer barrier, at
the same time pulls the system into a more chaotic transition state.

In Ref.\ \cite{gumn}, the microscopic origin of the instability against
asymmetric deformations was linked to specific quantum states having
their probability maxima in the central equatorial plane or in two
planes parallel to it. These states were energetically most sensitive
to the asymmetric deformations, whereas all others were practically
not. In order to check this behaviour in our present cavity model and
to investigate the relation of these states to the leading periodic
orbits, we have calculated its quantum spectrum (with Dirichlet
boundary conditions) at the relevant deformations. In \fig{lev4} we
show the wave number spectrum $k_i=\sqrt{2mE_i}/\hbar$ (in units of
$1/R$) for the lowest states $i$ with angular momentum $L_z=4$ obtained
at $c=1.53$ versus the asymmetry parameter $\alpha$. Whereas most
levels vary only little with $\alpha$, we clearly recognize a sequence
of states whose energies decrease more and more steeply with increasing
number $i$. These states are diabatic in that they connect different
portions of the level spectrum through avoided crossings; in the upper
part of the spectrum we have emphasized them by heavy grey lines and
labeled them by the numbers $i$ of the adiabatic levels counted from
the bottom. We shall show below that the energies of these diabatic
states can be obtained from an EBK quantization of the classical motion
near the equatorial planes.

The mechanism of the energy instability against $\alpha$ now

\Figurebb{lev4}{70}{35}{528}{720}{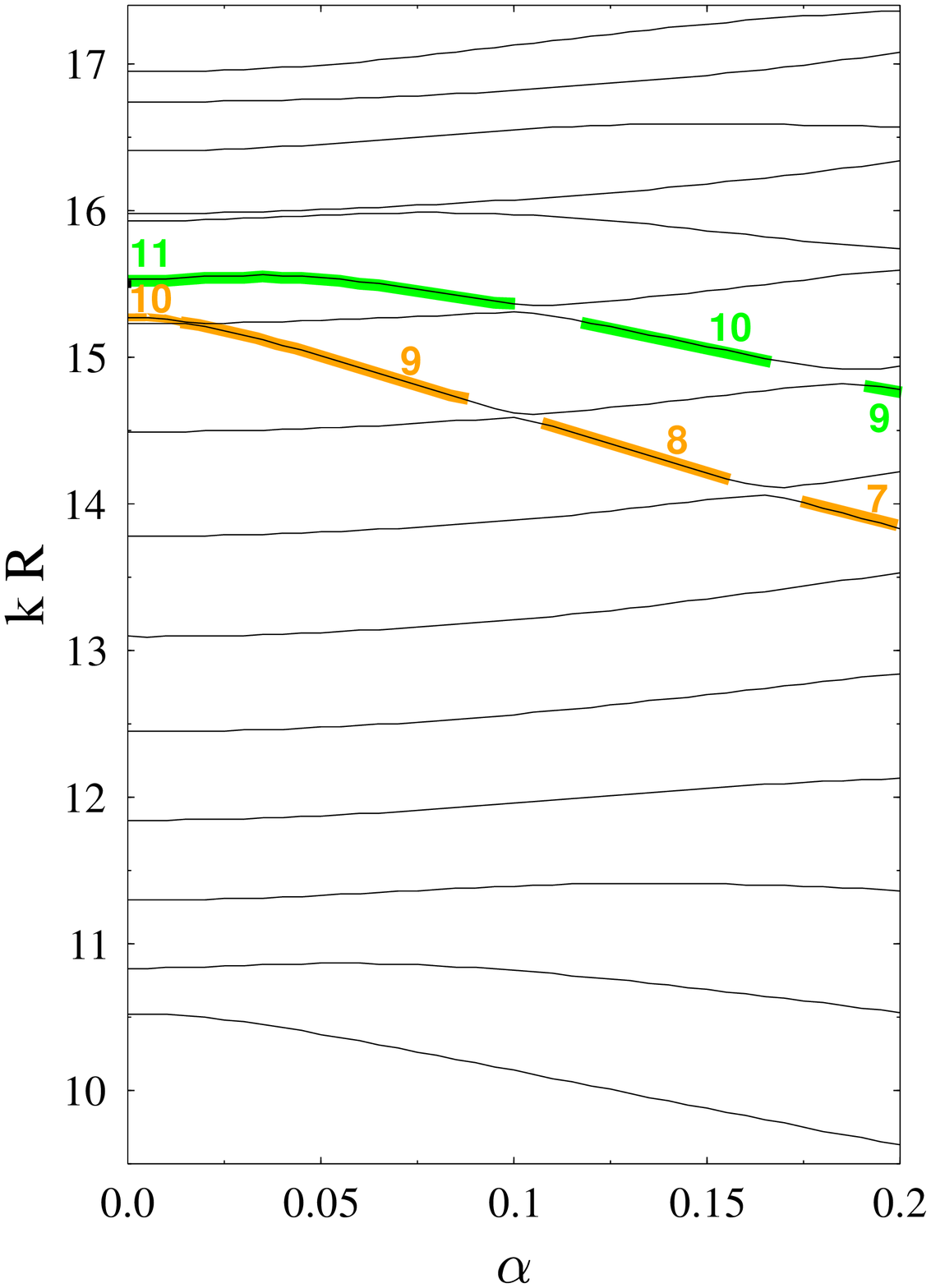}{9.0}{6.0}{
Quantum levels $k_iR$ for the lowest states $i$ with $L_z=4$ at
elongation $c=1.53$, plotted versus asymmetry $\alpha$. The emphasized
portions connected by avoided crossings denote some of the diabatic
states (see text).
}

\noindent
is this \cite{gumn}: if one or two of the diabatic states are located
just underneath the Fermi energy (and thus occupied), the system gains
energy when $\alpha$ is increased from zero. Since all the occupied
``inert'' states slightly increase at larger values of $\alpha$, the
sum of occupied levels (and hence $\delta E$, see Ref.\ \cite{stru})
will exhibit a minimum at some finite value $\alpha_0$. In our model
calculation \cite{fis1,fis2} for Pu$^{240}$ with a fixed Fermi energy
$k_F=12.1/R$, we obtained $\alpha_0 \simeq 0.13$.

In \fig{wfs} we show probability distributions by contour plots of the
squares of wave functions $|\psi_i(z,\rho)|^2$ in the $(z,\rho)$ plane.
The centre and bottom panels correspond to $\,i=10$ at $\alpha=0\,$ and
to $\,i=7$ at $\alpha=0.2$, respectively, representing one diabatic
state at the two ends of the $\alpha$ interval. The top panel
represents the  beginning of the next higher diabatic state, starting
as $i=11$ at $\alpha=0$. The probability maxima are clearly located
precisely in the planes $z_i$ of the shortest periodic orbits,
indicated by the heavy vertical lines. This pattern was found to be
consistent: the probability maxima of all diabatic states are located
in the planes of the shortest periodic orbits, whereas the maxima of
all other ``inert'' states could not be correlated to any of the
leading periodic orbits.

In order to quantify the correspondence between the diabatic quantum
states and the leading periodic orbits, we apply a method analogous to
the quasimode construc-

\Figurebb{wfs}{83}{30}{485}{760}{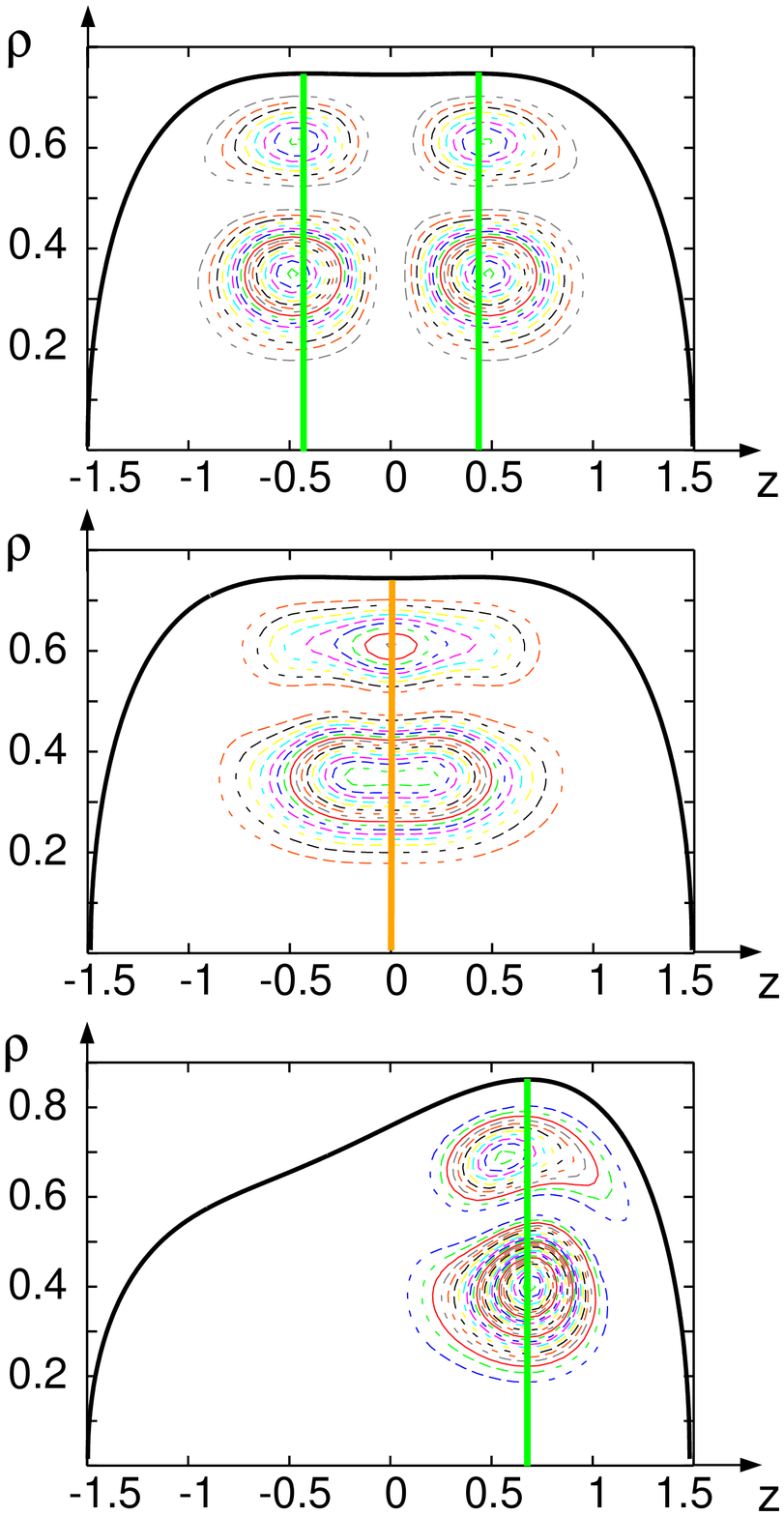}{10.0}{6.0}{
Contour plots of the probability distributions $|\psi_i(z,\rho)|^2$ for
selected states with $L_z=4$ at the elongation $c=1.53$. The heavy black
lines show the shape $\rho(z)$ of the boundary. The grey vertical lines
indicate the planes $z_i$ containing the shortest periodic orbits. {\it
Top:} state $i=11$ at asymmetry $\alpha=0.0$. {\it Centre:}  state
$i=10$ at asymmetry $\alpha=0.0$. {\it Bottom:}  state $i= 7$ at
asymmetry $\alpha=0.2$.
}

\noindent
tion for eigenfunctions located in the vicinity of stable periodic
orbits (see, e.g., \cite{voro}). Consider the classical motion in the
vicinity of a stable equatorial plane. It consists of small oscillations
around this plane. In a linearized approximation, the motion is
integrable and restricted to tori in phase space. The quasimode
approximation consists in applying the EBK quantization conditions to
this torus
structure \cite{rmrk}. To derive the quantization conditions, we use
the fact that the quasimode approximation depends only on the two
curvature radii at the boundary of the stable orbit plane: the radius
$R_1$ in the equatorial plane itself, and the radius $R_2$ perpendicular
to it. We therefore replace our billiard system locally by an axially
symmetric ellipsoid with the same curvature radii of the equatorial
planes. For an ellipsoid, the classical action variables are known
\cite{elli}. In the vicinity of the equatorial plane they are given by
the $z$ component of the angular momentum, $L_z/\hbar = 0,\pm1,\pm2,
\dots\,$, by the action $I_\rho$ for the radial motion in the
equatorial plane,
\bea
I_\rho & \approx & \frac{1}{\pi}\left[ \sqrt{p^2R_1^2-L_z^2}
         - |L_z| \arccos |L_z/(pR_1)| \right.
           \nonumber\\
       &   & \left.  - \,\frac{\kappa}{2} \,
             \frac{\arcsin\sqrt{R_1/R_2-L_z^2/(p^2R_1R_2)}}
                  {\sqrt{p^2 R_1(R_2-R_1) + L_z^2}} \right],
\label{irada}
\eea
and by the action $I_z$ for the motion out of this plane,
\be
I_z \approx \kappa/[2\sqrt{p^2 R_1(R_2-R_1) + L_z^2}]\,.
\label{iza}
\ee
These actions depend on the three constants of motion: the energy $E$
(or the momentum squared $p^2=\hbar^2k^2$), the angular momentum $L_z$,
and a conserved quantity $\kappa$ which vanishes for the motion in the
equatorial plane and is the analogue of the second constant of motion
in a two-dimensional elliptic billiard (given by the product of angular
momenta with respect to the two foci). The actions \eq{irada}, \eq{iza}
are already given here in the linearized approximation obtained from
the exact actions (given by elliptic integrals) through Taylor
expansion up to first order in $\kappa$. Eliminating $\kappa$ from
Eqs.\ \eq{irada}, \eq{iza}, we arrive at
\bea
I_\rho & \,= & \frac{1}{\pi}\left[ \sqrt{p^2R_1^2-L_z^2} -
             |L_z| \arccos |L_z/(pR_1)| \right. \nonumber\\
       &   & - \left. I_z\arcsin\sqrt{R_1/R_2-L_z^2/(p^2R_1R_2)}
               \right],
\label{irad}
\eea
to be used with the EBK quantization conditions
\bea
  I_z & = \,\hbar\,(n - 1/2)\,, \qquad & n = 1,2,3,\dots \label{ebkn} \\
  I_\rho &\,=\,\hbar\,(m - 1/4)\,. \qquad & m = 1,2,3,\dots \label{ebkm}
\eea
Note that if one puts $I_z=0$ above, Eqs.\ \eq{irad} and \eq{ebkm}
yield precisely the radial EBK quantization condition (without further
approximation) for a two-dimensional circular billiard with radius
$R_1$.

Equations (\ref{irad}) - \eq{ebkm} represent our quasimode
approximation. They implicitly yield the energies $E_{nm}$ (or wave
numbers $k_{nm}$) of quantum states with angular momentum $L_z$ located
near the planes of stable periodic orbits, which we expect to represent
the diabatic states discussed above. The quantum numbers $n$ and $m$
count the wave function extrema in the $z$ and $\rho$ directions,
respectively. (Correspondingly, $n$$-$$1$ and $m$$-$$1$ count the
numbers of nodes in the respective directions.) E.g., the squared
wavefunctions shown in the centre and bottom panels of \fig{wfs} should
represent the state $(1,2)$, and the one shown in the top panel should
belong to the state $(2,2)$.

As a test of our interpretation of the diabatic quantum states and
their semiclassical quantization, we compare in \fig{lev0} the exact
quantum spectrum $k_i$ (shown by solid lines) to the approximate EBK
levels $k_{nm}$, calculated here for the lowest states with angular
momentum $L_z=0$. Due to our linearization of the actions \eq{irada},
\eq{iza}, the agreement should be best for small values of the quantum
number $n$. It is, indeed, perfect for all states with $n=1$ shown in
\fig{lev0} by the dotted lines, which agree exactly with the quantum
levels of the corresponding diabatic states. The agreement is less good
for the states with $n=2$ (shown by the dashed lines), but our
semiclassical approximation still reproduces their correct
slopes for larger $\alpha$ and allows for their unique assignment.
Similar results are also obtained for larger values of $L_z$ (for which
the numerical agreement actually improves).

In summary, we have established a correspondence between the shortest
classical periodic orbits and a set of

\Figurebb{lev0}{60}{35}{560}{590}{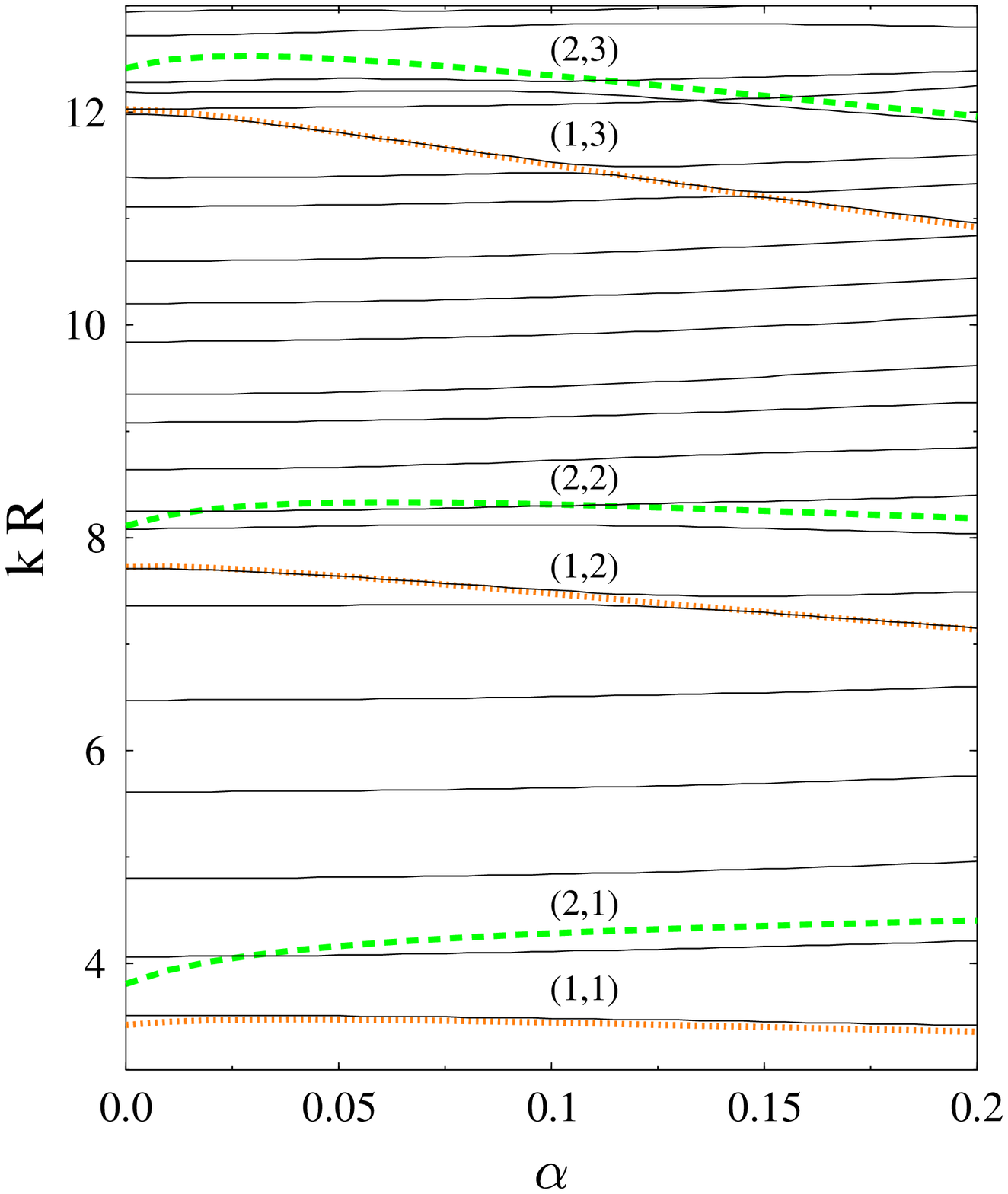}{9.5}{6.0}{
Quantum levels $k_iR$ of lowest states $i$ with $L_z=0$ at elongation
$c=1.53$ versus asymmetry $\alpha$ (solid lines). Labels $(n,m)$
indicate diabatic states; their levels $k_{nm}R$ obtained by the
quasimode approximation, Eqs.\ (\protect\ref{irad}) -
(\protect\ref{ebkm}), are shown by dotted lines for $n=1$ and dashed
lines for $n=2$.
}

\noindent
dia\-batic quantum states responsible for the onset of the
left-right asymmetry in a simple
axially deformed cavity model describing schematically the shapes of an
actinide nucleus on its adiabatic path to fission. The energy gain due
to the asymmetric deformations near the outer barrier had earlier been
reproduced by a semiclassical trace formula \cite{fis1,fis2}; thereby
the fission path through the deformation space was shown to be
determined by the constancy of the actions of the shortest periodic
orbits. In the present paper we have demonstrated that these orbits lie
at the centres of small islands of stability in an otherwise chaotic
phase space. Its degree of chaoticity is increased by the shell effect
causing the asymmetric deformations. We have also shown that the
diabatic quantum states which energetically favor the asymmetry
have their probability maxima precisely in the planes where the
shortest periodic orbits are located. A quasimode approximation based
on EBK quantization of the linearized classical motion around the
planes of the shortest stable orbits allowed us to uniquely assign
quantum numbers ($L_z,n,m$) to the diabatic quantum states and to
semiclassically reproduce their energies rather well.

\bs

We acknowledge the help of P. Meier in constructing the energy level
plots. This work has been supported by the Deutsche
Forschungsgemeinschaft (M.B.) and a Habilitationsstipendium des
Freistaates Bayern (S.M.R.).


\vspace{-0.2cm}

\begin{references}
\vspace*{-1.4cm}

\bibitem{fuhi} M. Brack, J. Damg{\aa}rd, A. S. Jensen, H. C. Pauli,
               V. M. Strutinsky, and C. Y. Wong, Rev.\ Mod.\ Phys.\
               {\bf 44}, 320 (1972).

\bibitem{sven} S. Bj{\o}rnholm and J. E. Lynn, Rev.\ Mod.\ Phys.\
               {\bf 52}, 725 (1980).

\bibitem{fis1} M. Brack, S. M. Reimann, and M. Sieber, Phys.\ Rev.\
               Lett.\ {\bf 79}, 1817 (1997).

\bibitem{fis2} M. Brack, P. Meier, S. M. Reimann, and M. Sieber,
               in {\it Similarities and Differences between Atomic
               Nuclei and Clusters}, eds.\ Y. Abe {\it et al.}
               (AIP New York, 1998), p.\ 17.

\bibitem{gutz} M. C. Gutzwiller, J. Math.\ Phys.\ {\bf 12}, 343 (1971).

\bibitem{bb}   R. Balian and C. Bloch, Ann.\ Phys.\ (N. Y.) {\bf 69},
               76 (1972).

\bibitem{stru} V. M. Strutinsky, Nucl. Phys. {\bf A 95}, 420 (1967);
               {\bf A 122}, 1 (1968).

\bibitem{asym} P. M\"oller and S. G. Nilsson, Phys.\ Lett.\ {\bf 31
               B}, 283 (1970); H. C. Pauli, T. Ledergerber, and M.
               Brack, Phys.\ Lett.\ {\bf 34 B}, 264 (1971).

\bibitem{gumn} C. Gustafsson, P. M\"oller, and S. G. Nilsson, Phys.\
               Lett.\ {\bf 34 B}, 349 (1971).

\bibitem{voro} A. Voros, Colloques Internationaux du CNRS no.\
               237 (Aix-en-Provence, France, 24-28 June 1974), ed.\
               J. M. Sourian (1975), p.\ 277.

\bibitem{rmrk} The quasimode treatment is limited to stable orbits.
               Note, however, that some wavefunctions are also found
               along planes of unstable orbits (see center of
               \fig{wfs}). This is similar to the scar phenomenon of
               the localization of wavefunctions near isolated
               unstable periodic orbits.
		
\bibitem{elli} P. H. Richter, A. Wittek, M. P. Kharlamov, and A. P.
               Kharlamov, Z. Naturforsch.\ {\bf 50a}, 693 (1995).

\end{references}
\end{document}